\documentstyle[preprint,aps]{revtex}
\tighten
\input{epsf.sty}
\font\blackboard=msbm10
\font\blackboards=msbm7
\font\blackboardss=msbm5
\newfam\black \textfont\black=\blackboard
\scriptfont\black=\blackboards
\scriptscriptfont\black=\blackboardss
\def\Bbb#1{{\fam\black\relax#1}}
\def\be{\begin{equation}}
\def\ee{\end{equation}}
\def\baray{\begin{eqnarray}}
\def\earay{\end{eqnarray}}

\def\N5{{1\over 2\times 5!}}

\def\Ecal{{\cal E}}

\def\dprime{{\prime\prime}}

\begin{document}

\title{Brane world models with bulk scalar fields}
\author{\'Eanna \'E. Flanagan\footnote{eef3@cornell.edu}, 
S.-H. Henry Tye\footnote{tye@mail.lns.cornell.edu} and Ira
Wasserman\footnote{ira@astro.cornell.edu}}
\address{Laboratory for Nuclear Studies and
Center for Radiophysics and Space Research \\
Cornell University \\
Ithaca, NY 14853}

\medskip

\date{\today}

\maketitle

\begin{abstract}

We examine several different types of five dimensional stationary spacetimes
with bulk scalar fields and parallel 3-branes. We study different methods for
avoiding  the appearance of spacetime singularities in the bulk for
models with and without cosmological expansion. For non-expanding models, we
demonstrate that in general the Randall-Sundrum warp factor is
recovered in the asymptotic bulk region, although elsewhere the
warping may be steeper than exponential. We show that
nonsingular expanding  models can be constructed as long as the
gradient of the bulk scalar field vanishes at zeros of the
warp factor, which are then analogous to the particle horizons
found in expanding models with a pure AdS bulk. Since the
branes in these models are stabilized by bulk scalar fields,
we expect there to be no linearly unstable radion modes. As an
application, we find a specific class of expanding, stationary
solutions with no singularities in the bulk in which the
four dimensional cosmological constant and mass hierarchy are
naturally very small.

\end{abstract}

\section{Introduction}

The Randall-Sundrum (RS) model\cite{RS1} for a warped 5-D geometry can 
account for the mass hierarchy, while providing a fine-tuning mechanism 
for canceling the 4-D cosmological constant without requiring a vanishing 
5-D vacuum energy\cite{RS1,RS2}. 
Variants of the model have been constructed in which
the cosmological constant is exponentially small\cite{TW}, and
in which both the cosmological constant and the hierarchy problems may be 
solved simultaneously\cite{FJSTW}.  However, the RS model and similar
models that solve the hierarchy problem are untenable: 
the radion-mediated interaction of matter on the visible brane dominates
over the gravitational interaction \cite{gold,garriga}.  This has the
consequence that matter on the visible brane (with positive brane tension) 
gives a negative
contribution to the square of the four dimensional Hubble parameter
for homogeneous cosmological models \cite{STW,FJSTW}.  Another
manifestation of the problem is that the spacetimes are dynamically
unstable: the 3-branes which they contain, whose positions must be
arranged carefully to reproduce the mass hierarchy, move away from
those special positions when slightly perturbed \cite{instability,FHJTW}. 

As suggested by Goldberger and Wise (GW)\cite{gold}, the introduction 
of bulk scalar fields can solve these problems by giving a mass to the 
radion mode and stabilizing the positions of the branes. This is 
consistent with phenomenology if the radion mass is at or above the 
electroweak scale\cite{Csaki}. DeWolfe, Freedman, Gubser and Karch 
(DFGK) \cite{DeWolfe} showed how fully consistent stationary spacetimes 
could be constructed with bulk scalar fields, including
the gravitational back-reaction of those fields; there have been
further studies of spacetimes with bulk scalars along similar
lines\cite{Csaki1,Zura}. DFGK argued that the
requirement that the induced metric on the brane be flat requires one
fine-tuning of the parameters describing the solutions.
Generalizing the class of solutions by
allowing a non-zero effective 4-D cosmological constant $\Lambda_4$
relaxes this fine-tuning \cite{DeWolfe}.  

The purpose of this paper is to extend the work of DFGK \cite{DeWolfe} 
and Kakushadze\cite{Zura}, to discuss some general properties of 5-D 
spacetimes with a bulk scalar field, and to apply those properties to 
the solution of the cosmological constant and the hierarchy problems.
The main points of this paper are as follows:

\begin{itemize}

\item 
The principal problem associated with the introduction of a bulk scalar
field is that generic stationary solutions contain timelike curvature
singularities in the bulk at finite distances from the branes.  This
occurs both for static solutions ($\Lambda_4 = 0$) and for stationary
solutions with cosmological expansion ($\Lambda_4 \ne 0$).  
Such singularities have already been encountered in work on
self-tuning of the cosmological constant \cite{selftune}.
It is possible to avoid these singularities in two ways. 
One well-known way is to simply 
orbifold or otherwise compactify the fifth dimension in such a way
that the singularity is never encountered.  A second way, which is
perhaps preferable, is to carefully choose the scalar field potential
in such a way that the occurrence of singularities is prevented.  
In Sec.\ \ref{setup} below, we discuss a simple condition (first 
considered by Kakushadze\cite{Zura}) for
identifying such preferred potentials for static spacetimes, and
present several examples.  One interesting 
example is obtained by compactifying a 11-D spacetime down to 5-D
with a single radion field which acts as a bulk scalar, as well as a
7-form field strength in 11-D descends to a non-dynamical 5-form field
strength in 5-D that generates a potential for the scalar field (Sec.\
\ref{sec:dr} below). By adjusting the 5-form field strength, this
model can be rendered nonsingular.

\item In the RS model, the metric's ``warp factor'' falls off
exponentially as one moves away from the Planck brane, and this
exponential fall-off underlies the RS solution of the hierarchy problem.
In many models with scalar fields, the fall-off of the warp factor
is {\it faster} than exponential, which facilitates solving the hierarchy
problem (see the models in Secs.\ \ref{sec:even} and \ref{sec:odd}
below).  This point was 
mentioned in passing for a specific model by DFGK.

\item We show that allowing the four dimensional cosmological constant
$\Lambda_4$ to be non-zero exacerbates the tendency to form spacetime
singularities in the bulk.  In particular, models with bulk scalar
fields that are nonsingular when $\Lambda_4=0$ should be expected to
become singular for nonzero $\Lambda_4$, although the corresponding
singularities are rather mild. The singularities are weak enough
(see Eq.~[\ref{ricci3}]) to be removed by simply requiring the bulk to
become effectively AdS as the warp factor $A(y)\to 0$.
In Sec.\ \ref{hnzero} below we derive a general method of
constructing nonsingular models with cosmological expansion based
on this idea.

\item We can now construct models without bulk curvature singularities.
With us living on the visible brane, it is possible to account for the 
extreme smallness of the electroweak scale and the cosmological 
constant simultaneously (Sec.\ref{sec:problems} below).  
These models are analogous to those of Refs.\ \cite{TW,FJSTW}, 
with the additional feature that the stability problems of Refs.\
\cite{TW,FJSTW} have been cured.

\end{itemize}

\section{Setup and General Considerations}
\label{setup}

In this section we outline the general framework, review the
procedure introduced by DFGK for constructing static and stationary
solutions, and derive criterion under which bulk singularities do not
occur.

\subsection{Basic equations}

We consider 5-D gravity plus a bulk scalar field with 
parallel 3-branes, for which the action is
\be
\label{action}
 S=\int d^4xdy \sqrt{\vert g\vert}\left[\frac{1}{2\kappa^2}R -
\frac{(\nabla \phi)^2}{2} - V(\phi) \right]
- \sum_b \int_{y_b} d^4x \sqrt{\vert \tilde g^b \vert} \sigma_b(\phi).
\ee
Here the coordinates are $(x^\mu,y)$ for $0 \le \mu \le 3$, 
the $b$th brane is located at $y=y_b$, $g_{ab}$ is the 5-D metric and 
$\tilde g^b_{\mu \nu}$ is the induced metric on the $b$th brane.
The brane tensions $\sigma_b$ and potential $V$ are functions of the
bulk scalar $\phi$, and $\kappa^2$ is the 5-D gravitational 
coupling constant.

We seek solutions to the field equations of the form
\begin{eqnarray}
ds^2&=&dy^2+A(y)\left[-dt^2+e^{2Ht}\delta_{ij}dx^idx^j\right], \\
\phi &=& \phi(y),
\label{metric}
\end{eqnarray}
although we shall concentrate initially on the static case $H=0$.
For this ansatz, the Einstein
and scalar field equations reduce to \cite{DeWolfe}
\baray
\label{keyeq}
u^\prime&=&-{2 \kappa^2 (\phi^\prime)^2\over
3}- {2H^2\over A} -
{2\kappa^2\over 3}\sum_b\sigma_b (\phi)
{\delta(y-y_b)}, \nonumber\\
u^2&=&{\kappa^2(\phi^\prime)^2\over 3}-{2\kappa^2 V(\phi)\over 3} 
+{4H^2\over A}, \nonumber\\
\phi^\dprime + 2u\phi^\prime &=& {\partial V(\phi)\over\partial \phi}
+ \sum_b \left[{\partial \sigma_b\over\partial\phi}\right]\delta(y-y_b),
\earay
where 
\be
u(y)= {A^\prime \over A}
\label{udef0}
\ee
and primes denote derivatives with respect to $y$.
Only two of these three equations are independent:
the scalar wave equation follows from the other two via the Bianchi
identities.  
The jump conditions at the branes are
\be
u\big\vert^{y_b^+}_{y_b^-} \equiv
\lim_{y\to y_b +}u- \lim_{y\to y_b -}u = -2q_b, \quad \quad
\phi^\prime\big\vert^{y_b^+}_{y_b^-}  
= \sigma_b^\prime(\phi_b),
\label{Israel}
\ee
where $\phi_b = \phi(y_b)$ is the value of the bulk scalar on the $b$th
brane and 
\be
q_b = { 1 \over 3} \kappa^2 \sigma_b(\phi_b).
\ee
The Ricci scalar is
\begin{eqnarray}
\label{ricci0}
R &=& -4u^\prime-5u^2+{12H^2\over A} \\
 &=& \kappa^2(\phi^\prime)^2+{10\kappa^2V(\phi)\over 3}.
\label{ricci}
\end{eqnarray}

\subsection{Method of obtaining solutions}

We now review the method of generating solutions introduced by DFGK.  
For any given solution of Eqs.\ (\ref{keyeq}), we can imagine
inverting the relation $\phi = \phi(y)$ to obtain $y$ as a function of
$\phi$.  Leaving aside questions of single-valuedness, we can imagine  
changing independent variables from $y$ to $\phi$. Since
$\phi^\prime(y)$ is also a function of $y$,
we can likewise take $\phi^\prime$ to be a function
of $\phi$. By analogy with the practice in supergravity theory, let
us define a function $W(\phi)$ by
\be
\phi^\prime\equiv{1\over 2}{\partial W(\phi)\over\partial\phi}.
\label{phipsuper}
\ee
In the $H=0$ case, the first of Eqs.\ (\ref{keyeq}) then
implies that, away from branes,
\be
{\partial u\over\partial\phi}=-{2\kappa^2\over 3}\phi^\prime=-
{\kappa^2\over 3}{\partial W(\phi)\over\partial\phi},
\ee
so that 
\be
u=-\kappa^2W(\phi)/3.
\label{udef}
\ee
The second
of Eqs.\ (\ref{keyeq}) then becomes
\be
V(\phi)={1\over 8}\left[{\partial W(\phi)\over\partial\phi}\right]^2
-{1\over 6} \kappa^2 W(\phi)^2.
\label{vsuper}
\ee
The procedure for obtaining solutions in the $H=0$ case is (i) choose
a potential $V(\phi)$ and superpotential $W(\phi)$ that are related by
Eq.\ (\ref{vsuper}); (ii) integrate Eq.\ (\ref{phipsuper}) to obtain
$\phi$ as a function of $y$; (iii) combine this with Eq.\ (\ref{udef})
to obtain $u$ as a function of $y$; and (iv) integrate Eq.\ (\ref{udef0})
to obtain the warp factor $A$ as a function of $y$.

\subsection{Occurrence of singularities in the bulk}
\label{sec:avoid}

Combining the definition (\ref{phipsuper}) of the superpotential $W$ with the
expression (\ref{ricci}) for the Ricci scalar gives
\be
R={\kappa^2\over 4}\left[{\partial W(\phi)\over\partial\phi}\right]^2
+{10\kappa^2V(\phi)\over 3}.
\label{ricciphi}
\ee
Eq.~(\ref{ricciphi}) is applicable to models with $H \ne 0$.
When $H=0$, substituting Eq.\ (\ref{vsuper}) into Eq.\
(\ref{ricciphi}) yields
\be
R={2\kappa^2\over 3}\left[{\partial W(\phi)\over\partial\phi}\right]^2
-{5\kappa^4 W^2(\phi)\over 9}.
\label{ricciphinoexp}
\ee
Eqs.~(\ref{ricciphi}) and (\ref{ricciphinoexp}) show why
singularities are rampant in models with bulk scalar fields. For many
choices of $W(\phi)$ and $V(\phi)$, it turns out that $\vert R(\phi)\vert
\to\infty$ as $\vert\phi\vert\to\infty$.  Examples include
$V(\phi)=\mu^2\phi^2/2$ or $V(\phi)=\lambda\phi^4/4$ or $V(\phi)
=V_0e^{-\phi/\phi_0}$, all of which are simple, and even
well-motivated in some respects. Thus, unless $\phi$ is
constrained to remain finite everywhere, a singularity will be
encountered. 

To see how to avoid singularities in general\cite{Zura}, consider
first a spacetime with a single brane, at $y=y_b$, and suppose that
the value of the scalar field on that brane is $\phi_b$.  From Eq.\
(\ref{phipsuper}), we can solve for $y$ as a function of $\phi$:
\be
y(\phi) = y_b + I(\phi_b,\phi),
\ee
where
\be
I(\phi_b,\phi)\equiv\int_{\phi_b}^{\phi}{d\phi\over\partial
W(\phi)/\partial\phi}.
\label{Idef}
\ee
Now if there exist finite values $\phi_{-\infty}$ and $\phi_{+\infty}$
of the bulk scalar field such that $\vert
I(\phi_b,\phi_{\pm\infty})\vert =\infty$,
then the spacetime can be infinite in the fifth dimension with 
$\phi\to\phi_{\pm\infty}$ as $y \to -\infty$ and $y \to \infty$.
In such cases, an infinite range in $y$
is mapped onto a finite range in $\phi$, thus potentially avoiding any
singularities. 

From Eq.\ (\ref{Idef}), divergences of $I(\phi_b,\phi)$ will occur
only at zeros of $W^\prime(\phi)$.  Suppose now that $W^\prime(\phi)$
has a zeros at both $\phi > \phi_b$ and $\phi < \phi_b$, and that
$W^\prime(\phi)$ vanishes at those zeros at least linearly (but
not like $\vert\phi - \phi_{\pm \infty}\vert^p$ with $p < 1$).
If, in addition, neither
$W^\prime(\phi)$ nor $V(\phi)$ diverges in between
these zeros, then the spacetime can extend infinitely in the fifth
dimension away from the brane in either direction, without 
any singularities.  A similar result can be obtained for spacetimes
with more than one brane.  In this case, the condition is that 
$I(\phi_b, \phi_{b^\prime})$ be finite for any
two branes $b,b^\prime$, and that for the two bounding branes, $b_\pm$,
there exist $\phi_{\pm\infty}$ such that
$I(\phi_{b_\pm},\phi_{\pm\infty})=\pm\infty$ with no intervening
infinities in $V(\phi)$ and $\partial W(\phi)/\partial\phi$.

In models where these conditions for avoiding singularities are
met, $V(\phi)$ tends to a constant asymptotically, as a zero is
approached. Thus, such models, for $H=0$, tend toward the RS
solution far from branes, provided that the scalar field
potential is negative. However, there will also be regions in such
solutions where the spacetime metric differs considerably from
the RS model, and where the warp factor drops far faster than
exponentially.  We shall show how this comes about in a 
specific model below. The $H\neq 0$ solutions may also tend
asymptotically toward their counterparts with uniform bulk
cosmological constant \cite{FJSTW}.  However, 
the relatively harmless particle horizons characteristic 
of those uniform bulk cosmological constant models will generically be
transformed in to curvature singularities, unless there are fortuitous
cancellations, as can be seen from the $A \to 0$ limit of Eq.\
(\ref{ricci0}).

Models where singularities would be inevitable in an uncompactified
geometry can be truncated by compactification to a region that does
not include any singularities. Such models need not approach 
the RS model, even asymptotically, and can have warp factors
that vary much more rapidly than exponentially throughout the compactified
fifth dimension. 

\section{Models without cosmological expansion}
\label{hzero}

In this section we discuss several different choices of superpotential
$W(\phi)$ and the properties of the corresponding solutions in the
static case $H=0$.

\subsection{Even superpotential}
\label{sec:even}

The RS model corresponds to $W(\phi)=\pm a$, where $a>0$
is a constant; this model has $V=-{\kappa^2a^2/6}$.
DFGK showed that the GW stabilization
mechanism can be implemented if instead $W(\phi)=\pm(a-b\phi^2)$, where 
$b>0$ is another constant, for which
\be
V(\phi)=-{\kappa^2a^2\over 6}+{\phi^2\over 2}\left(b^2
-{2\kappa^2ab\over 3}\right)-{\kappa^2b^2\phi^4\over 6}.
\label{gw1}
\ee
Choosing the $+$ sign in $W(\phi)$ for $y>0$ (and vice-versa), DFGK
found $\phi(y)=\phi(0)e^{-by}$, and therefore (via Eq.~[\ref{keyeq}])
$3u/\kappa^2=-a+b\phi^2(0)e^{-2by}$, which implies a transition
between two regions 
of constant $u$, one at small $by$ and the other at large
$by$, where the model becomes equivalent to the RS model. This model
has $\partial W(\phi)/\partial\phi=0$ at $\phi=0$, and is nonsingular as
long as the maximum value of $|\phi|$ is finite. Thus, it avoids
singularities by mapping an infinite range of $y$ to a finite range
in $\phi$, as was discussed in Sec.\ \ref{sec:avoid}.

DFGK noted that they could also choose $W(\phi)=\pm(a+b\phi^2)$,
with $b>0$, for which
\be
V(\phi)=-{\kappa^2a^2\over 6}+{\phi^2\over 2}\left(b^2
+{2\kappa^2ab\over 3}\right)-{\kappa^2b^2\phi^4\over 6}.
\label{gw2}
\ee
Eqs.~(\ref{gw1}) and (\ref{gw2}) can be rewritten in the form
\be
V(\phi)=-\Lambda_b+{\mu^2\phi^2\over 2}-{\lambda_\mp\phi^4\over 4},
\label{gw}
\ee
where the $\mp$ signs apply to Eqs.(\ref{gw1}) and (\ref{gw2})
respectively, and where
\be
\lambda_\mp={2\kappa^2\over 3}\left(\sqrt{\mu^2+{2\kappa^2\Lambda_b\over 3}}
\mp\sqrt{2\kappa^2\Lambda_b\over 3}\right)^2.
\ee
Despite appearances, the differences between the two models 
(\ref{gw1}) and (\ref{gw2}) are not minor. 
For the $+$ model, we have $W(\phi)=a 
+b\phi^2$ for $y>0$, so that $\phi^\prime=b\phi$,
$\phi(y)=\phi(0)e^{by}$ and $3u/\kappa^2=-a-b\phi^2(0)e^{2by}$.
This yields a warp factor $A(y)$ which is an exponential of an
exponential:
\be
\ln A(y)=\ln A(0)-{1 \over 3} \kappa^2ay- {1 \over 6}
\kappa^2 \phi(0)^2 \left[e^{2by}-1\right].
\ee
This super-warped version of the GW model, obtained
from an apparently insignificant change to the
potential (\ref{gw}), becomes singular if the fifth
dimension is not compactified, because $R(\phi) \to -\infty$ as $\phi
\to \infty$.
Any compactification of the fifth dimension avoids the singularity
by construction. The model can be altered slightly to become bulletproof
against singularities by taking  
$W(\phi)=\pm(a+b\phi^2-b\phi^4/2\phi_\infty^2)$.  For this choice of
$W$ we have
\be
\phi(y)={\phi(0)\phi_\infty e^{by}\over\sqrt{\phi^2_\infty-\phi^2(0)
+\phi^2(0)e^{2by}}},
\ee
so that $\phi\to\phi_\infty$
and $u\to -(\kappa^2/3)(a+b\phi_\infty^2/2)$ at large $y$, 
reducing asymptotically to the RS model.
Note that the total change in the logarithm of the
warp factor in the super-warped regime is limited
to $\sim\kappa^2\phi^2_\infty$, which may be large.

\subsection{Odd superpotential and Gaussian warp factor}
\label{sec:odd}

The DFGK realizations of the GW stabilization
mechanism for the RS model are based on choices of superpotential
$W(\phi)$ that are even functions of $\phi$. There are also simple
models in which $W(\phi)$ is odd, first considered in Ref\cite{Zura}; 
the simplest is 
\be
W(\phi)=\pm 2b\phi,
\label{odd1}
\ee
with $b>0$.
(A constant term in $W(\phi)$ can be absorbed into the definition of
$\phi$.)  For the model (\ref{odd1}), we find that 
\baray
\phi&=&\phi(0)+by,\nonumber\\
u&=&-{2\kappa^2b\phi(0)\over 3}-{2\kappa^2b^2y\over 3},\nonumber\\
\ln A(y)&=&\ln A(0)-{2\kappa^2b\phi(0)y\over 3}-{\kappa^2b^2y^2\over 3},
\earay
for $y>0$. Thus, the warp factor in this model is Gaussian rather than
exponential, and therefore falls off more rapidly than in the RS
model. Note that $A(y)$ has a maximum value, and all positive tension
branes must be located away from that maximum.

This model contains a singularity at $y \to \infty$ since $\phi(y)$
diverges there.  The singularity can be 
avoided by compactification, or can be removed altogether
by setting $W(\phi)=\pm 2b(\phi-\phi^3/3\phi_\infty^2)$,
in which case
\be
\phi(y)=\phi_\infty\tanh\left[\tanh^{-1}\left({\phi(0)\over
\phi_\infty}\right)
+{by\over\phi_\infty}\right]
\ee
which tends to $\phi_\infty$ asymptotically. The warp factor for this
model is given by
\baray
\ln A(y)&=&\ln A(0)-{4\kappa^2\phi_\infty^2\over 9}
\biggl\{\ln\left[\cosh\left({by\over\phi_\infty}\right)
+{\phi(0)\over\phi_\infty}
\sinh\left({by\over\phi_\infty}\right)\right]\nonumber\\& &
+{1\over 4}\left(1-{\phi^2(0)\over\phi_\infty^2}\right)
\left[1-{1\over\left[\cosh(by/\phi_\infty)
+\phi(0) \sinh(by/\phi_\infty) / \phi_\infty\right]^2}\right]
\biggr\},
\earay
and tends toward $\ln A(y)\simeq -4\kappa^2b\phi_\infty y/9$
for $by/\phi_\infty\gg 1$, which is the same scaling as in
the RS model. The maximum change in the logarithm of the
warp factor in the Gaussian-warped regime is of order 
$\sim\kappa^2\phi_\infty^2$, similar to what we found for the
model of Sec.\ \ref{sec:even} above.

\subsection{Exponential potential}

The models considered so far become singular only in the asymptotic
regime $y \to \infty$.  Thus, whether or not these singularities
merit a potential-altering cure is largely a matter of taste,
since compactifying to any finite size in the fifth dimension
removes the singularity without any sort of fine-tuning.
However, there are other models which are ``spontaneously singular''
in the sense that they develop singularities at finite $y$. 
An example is when the superpotential is an exponential,
\be
W(\phi)=2ae^{-k\phi}.
\label{expmodel}
\ee
For this model, we have 
\be
e^{-k\phi}=\left[e^{k\phi(0)}-k^2ay\right]^{-1},
\ee
which diverges at $y=(k^2a)^{-1}e^{k\phi(0)}$, implying a divergence
of the Ricci scalar.  The singularity is removed by the following slight
modification of the superpotential,  
\be
W(\phi)=2a\left[e^{-k\phi}-{k\over q}e^{(q-k)\phi_\infty-q\phi}\right],
\label{expnonsing}
\ee
where $q>k$ is a constant.  The associated potential is
\baray
V(\phi)&=&{a^2e^{-2k\phi_\infty}\over 2}
\biggl[e^{-2k(\phi-\phi_\infty)}\left(k^2-{4\kappa^2\over 3}\right)
-2e^{-(k+q)(\phi-\phi_\infty)}\left(k^2-{4\kappa^2k\over 3q}\right)
\nonumber\\& &
+e^{-2q(\phi-\phi_\infty)}\left(k^2-{4\kappa^2k^2\over 3q^2}\right)\biggr].
\label{potexp}
\earay
For this model the relation between $\phi$ and $y$ is given by 
\be
\int_{e^{k(\phi-\phi_\infty)}}^{e^{k(\phi(0)-\phi_\infty)}}
{d\eta\,\eta^{q/k-1}\over\eta^{q/k-1}-1}=k^2ae^{-k\phi_\infty}y,
\label{phiint}
\ee
which can be integrated numerically in general.  The integral can be
done analytically in special cases.  For example, 
for $q=2k$, $q=3k$, and $q=5k$ we find the results
\baray
e^{k(\phi-\phi_\infty)}+\ln\left[e^{k(\phi-\phi_\infty)}-1\right]
=e^{k(\phi(0)-\phi_\infty)}
+\ln\left[e^{k(\phi(0)-\phi_\infty)}-1\right]-k^2ae^{-k\phi_\infty}y,
\nonumber\\
e^{k(\phi-\phi_\infty)}+{1\over 2}\ln\left[\tanh{k\over 2}
(\phi-\phi_\infty)\right]
=e^{k(\phi(0)-\phi_\infty)}+{1\over 2}\ln\left[\tanh{k\over 2}
(\phi(0)-\phi_\infty)\right]-k^2ae^{-k\phi_\infty}y,\nonumber
\earay
and
\baray
e^{k(\phi-\phi_\infty)}+{1\over 4}\ln\left[\tanh{k\over 2}
(\phi-\phi_\infty)\right]-{1\over 2}\tan^{-1}\left(e^{k(\phi-\phi_\infty)}
\right)=\nonumber\\
e^{k(\phi-\phi_\infty)}+{1\over 4}\ln\tanh\left[{k\over 2} 
(\phi-\phi_\infty)\right]-{1\over 2}\tan^{-1}\left(e^{k(\phi-\phi_\infty)} 
\right)-k^2ae^{-k\phi_\infty}y,
\earay
respectively.
For large values of $y$ this model reduces to the RS model, since 
$\phi\to\phi_\infty$ from above and $3u/\kappa^2\to
-2ae^{-k\phi_\infty}(1-k/q)$, a constant,

\subsubsection{Specific realization obtained from dimensional reduction}
\label{sec:dr}

A potential reminiscent of the potential (\ref{potexp}) arises from
compactification of an 11-D spacetime to 5-D; the result is
\be
V(\psi)=-\Lambda_be^{-2\psi/3}-c_6e^{-\psi}+{\Ecal^2\over 2}e^{-2\psi}.
\label{pot11}
\ee
Here it is assumed that the six compactified dimensions have a single
associated radion field, $\psi$, the curvature associated with the
compactified dimensions is $c_6>0$, $\Lambda_b$ is the descendant of
an 11-D cosmological constant, and $\Ecal$ is a 5-form field strength
(descended from a 7-form field strength in 11-D).  
Although the potentials (\ref{potexp}) and (\ref{pot11}) are similar,
there are no choices of parameter values for which the potentials
match up term by term.  This is because 
the exponents in three terms in Eq.\ (\ref{pot11}) are
in the ratios $1:1.5:3$, whereas 
the exponents in Eq.\ (\ref{potexp}) are in the ratio
$1:(k+q)/(2k):q/k$.  However, we can find parameter choices for which
the potentials match if we allow one of the two parameters 
$c_6$ and $\Lambda_b$ to vanish.  

If we drop the term proportional to $\Lambda_b$ and retain curvature
in the six compactified dimensions, we find two possible
identifications: 
\be
q=2k, \ \ \ \ \ k^2=2\kappa^2/3,\ \ \ \ \
\Ecal^2=c_6=(\kappa^2a^2/3)e^{-2k\phi_ \infty},
\label{comp1}
\ee
and
\be
q=3k, \ \ \ \ \ k^2=4\kappa^2/27, \ \ \ \ \Ecal^2=c_6=
(16\kappa^2a^2/27)e^{-2\phi_\infty}.  
\label{comp2}
\ee
If we set $c_6=0$ but retain
$\Lambda_b\neq 0$, there are also two possibilities:
\be q=3k, \ \ \ \ \ k^2=4\kappa^2/9, \ \ \ \ \
\Lambda_b=(4\kappa^2a^2/9)e^{-2k\phi_\infty},\ \ \ \ \
\Ecal^2=(8k^2a^2/27)e^{-2k\phi_\infty},
\label{comp3}
\ee
and
\be
q=5k,\ \ \ \ \ k^2=4\kappa^2/75, \ \ \ \ \
\Lambda_b=(12\kappa^2a^2/25)e^{-2k\phi_\infty}, \ \ \ \ \
\Ecal^2=(32\kappa^2a^2/75)e^{-2k\phi_\infty}.
\label{comp4}
\ee
Whichever possibility 
we choose, we can regard the combination $\kappa^2a^2e^{-2k\phi_\infty}$
as a derived quantity, determined by either $\Lambda_b$ or $c_6$,
depending on which is nonzero. The 5-form field strength $\Ecal$, which
involves a constant of integration, can then be adjusted to produce
a nonsingular model. The 5-form fields therefore may play a central role
in eliminating spacetime singularities from 5-D models derived by
dimensional reduction from 11-D. 

\section{Models with cosmological expansion}
\label{hnzero}

\subsection{Bulk singularities at zeros of the warp factor}

When the bulk geometry is anti-deSitter (AdS) space,
metrics of the form (\ref{metric}) have non-singular particle horizons
at zeros of the warp factor $A(y)$.  
From the expression (\ref{ricci0}) for the Ricci scalar, one might
anticipate a singularity from the term $\propto 1 / A$, but that
divergence is cancelled by divergences in the $u^2$ and $u^\prime$
terms for an AdS bulk.  

However, when scalar fields are included, the surfaces where $A(y)=0$
generically correspond to curvature singularities.
This is the case even when we set $\phi^\prime={1\over 2}\partial W(\phi)/
\partial\phi$ and choose $\partial W(\phi)/\partial\phi$ so that 
$\phi$ remains finite over the entire range of $y$.  The simplest way
to see this is to consider the first of Eqs.~(\ref{keyeq}) in a
neighborhood of a point where $A(y)=0$. Suppose that $\phi^\prime$ is
finite and nonzero 
in that neighborhood, so that as long as we restrict attention to a small
enough region, we can take it to be a constant. If $q^2\equiv
\kappa^2(\phi^\prime)^2/3$, then
\be
u^\prime=\left({A^\prime\over A}\right)^\prime=-2q^2-{2H^2\over A},
\ee
This equation would be
exact for $\partial W(\phi)/\partial\phi=$ constant, which led us
to the Gaussian model in Sec.\ \ref{hzero}. Multiply by $u=A^\prime/A$
and integrate to find
\be
\label{ricci3}
u^2=\left({A^\prime\over A}\right)^2=-4q^2\ln\vert A\vert
+{4H^2\over A}+4k^2.
\ee
Here $k^2$ is a constant which may be positive or negative, although 
we shall take it to be positive without exception to recover the
RS model when $q=0=H$.
Substituting these results for $u^\prime$ and $u^2$ into 
Eq.\ (\ref{ricci0}), we find
\be
R=-20k^2+8q^2+20q^2\ln\vert A\vert
\ee
as $A\to 0$. Thus, the Ricci scalar diverges at this point, which is
a true singularity, unless $q^2=0$. The divergence is relatively
mild (logarithmic). In fact, for $H=0$ this is the same divergence
as was found asymptotically in the Gaussian model of Sec.\ \ref{hzero}.

\subsection{Method of constructing singularity-free solutions}

In the singularity-free models with $H=0$, we saw that obtaining a mapping
of the infinite $y$ domain to a finite range of $\phi$ required that
$\phi^\prime\to 0$ asymptotically. For $H=0$ models the singularity is
absent, because $A\to 0$ only asymptotically, and 
$q^2\ln A\to 0$ asymptotically as well.  
For $H \ne 0$, nonsingular models can be constructed parametrically by
the following procedure.  Define a function $g(A)$ by
\be
{2\kappa^2(\phi^\prime)^2\over 3}=2A{dg(A)\over dA}\geq 0,
\label{phieq}
\ee
with $g(0)=0$.  
Then Eq.\ (\ref{ricci3}) generalizes to
\be
u^2=\left({A^\prime\over A}\right)^2
=-4g(A)+{4H^2\over A}+4k^2.
\label{ueq}
\ee
Comparing this with the second of Eqs.\ (\ref{keyeq}) we obtain
\be
{2\kappa^2V(\phi)\over 3}=A{dg(A)\over dA}
+4g(A)-4k^2.
\label{Veq}
\ee
The procedure for obtaining a solution can now be summarized as
follows:  (i) pick a function $g(A)$ and the integration constant
$k^2$;  (ii) solve Eq.\ (\ref{ueq}) to obtain $A$ as a function of
$y$;  (iii) solve Eq.\ (\ref{phieq}) to obtain $\phi$ as a function of
$y$;  (iv) use Eq.\ (\ref{Veq}) to compute the potential
$V(\phi)$ corresponding to the solution just obtained.
By construction, the Ricci
scalar
\be
R=-20k^2+8A{dg\over dA}+20g(A)
\ee
is finite at $A\to 0$.

Let us illustrate this procedure with the particular example
\be
g(A)=q^2A^p,
\label{gchoice}
\ee
with $p>0$.  (For small $p$, we expect this model to
resemble the singular Gaussian model.)  For the choice (\ref{gchoice})
we find 
\be
\left({A^\prime\over A}\right)^2=4k^2-4q^2A^p+{4H^2\over A}.
\label{peq}
\ee
Apparently, there is a maximum value of $A(y)$ for such models;
for small $H^2/k^2$ we find $A_{max}\simeq(k/q)^{2/p}+H^2/pk^2$.  
Note that in this particular model, freedom to rescale $A(y)$ also
implies that $q$ may be rescaled arbitrarily; this need not be
true for $g(A)$ that are not scale-free.
When $H=0$, the solution to  
Eq.\ (\ref{peq}) is (apart from an overall ambiguity in the sign
of $\phi^\prime$ and hence $\phi$)
\baray
\label{Aformula}
A(y)&=&\left({k\over q\cosh pky}\right)^{2/p}\\
\label{phiformula}
\phi\sqrt{\kappa^2p\over 12}
&=&\tan^{-1}\left(e^{pky}\right)-{\pi\over 4}\\
{2\kappa^2V(\phi)\over 3}&=&4k^2\left[-1+\left(1+{p\over 4}\right)
\cos^2\left(\phi\sqrt{\kappa^2p\over 3}\right)\right],
\label{psolhz}
\earay
where we have located the maximum of $A(y)$ at $y=0$ and chosen
$\phi=0$ at $y=0$.  As $y$ ranges from zero to infinity,
$\phi\sqrt{\kappa^2p/3}$ 
ranges from $0$ to $\pi/2$. 
Note that the maximum value of $V(\phi)$
is positive in this model, and occurs at the maximum of $A(y)$;
this is already apparent from Eqs.\ (\ref{ueq}) and (\ref{Veq}).
In a particular realization of this model, with branes at specific
values of $\phi$, the maximum value of the warp factor may
never be encountered. Finally, notice that for small values of
$p$, the warp factor $A(y)\sim e^{-pk^2y^2}$ for $pky\lesssim 1$,
and, for any value of $p$, $A(y)\sim e^{-ky}$ for $pky\gtrsim 1$.
These features are reminiscent of the modified nonsingular
Gaussian model of Sec.\ \ref{hzero}, and the potential
(\ref{psolhz}) therefore represents an alternative way of regularizing
the Gaussian model to avoid singularities.

When $H=0$, Eq.\ (\ref{Aformula}) shows that $A(y)\to 0$ only when
$y\to\infty$.  When $H\neq 0$, Eq.\ (\ref{peq}) shows that 
$\vert A^\prime/A\vert$ is larger than for $H=0$, and we therefore
expect the position of the zero of $A(y)$ to move inward from
$y=\infty$ to finite $y$.  If $y_0$ is the zero, so that $A(y_0)=0$, then
$A(y)$ may be found by inverting the equation
\be
2k(y_0-y)=\int_0^A{dA\over A\sqrt{1-A^p/k^2+H^2/k^2A}},
\label{Aeq}
\ee
and the scalar field may be found from
\be
\phi(A)=\phi(0)-\left({3p\over 4k^2}\right)^{1/2}
\int_0^A{dA\,A^{p/2-1}\over\sqrt{k^2-A^p+H^2/A}}.
\label{phiA}
\ee
Here $\phi(0)$ is the value at $A=0$, and we have assumed
$\phi^\prime>0$.  Eqs.\ (\ref{phiA}) and (\ref{Veq}) may be used
to evaluate the potential $V(\phi)$, which will have a nontrivial
dependence on $H$.
For $A\lesssim H^{2/(1+p)}$, the scalar field plays an 
insignificant role, and we recover the vacuum solution
$A(y)\simeq (H^2/k^2)\sinh^2[k(y-y_0)]$.   For $A\gtrsim
H^{2/(1+p)}$, on the other hand, we can neglect $H^2$, and we recover
the solution (\ref{Aformula}) for $A(y)$.
Corrections to these approximate solutions may be obtained from
Eq.\ (\ref{Aeq}). In general, the behavior of models with scalar
fields is subtle for small values of $H^2$, because nonsingular
models require that the effects of the scalar field become insignificant
as $A(y)\to 0$, so that the $H^2$ terms become important in
these regions despite the smallness of $H^2$.

These nonsingular models appear to share common features with
models developed previously without bulk scalar fields \cite{FJSTW},
particularly the occurrence of non-singular surfaces where $A(y)=0$.
Presumably, these surfaces are particle 
horizons similar to those which occur in the pure AdS case.

\section{Cosmological Constant and Hierarchy Problems}
\label{sec:problems}

To examine the implications of these models for the value of
the cosmological constant, focus again on the model (\ref{gchoice}) 
and rewrite Eq.\ (\ref{ueq}) as
\be
u^2=\left({A^\prime\over A}\right)^2=4\left(k^2
-{\kappa^2(\phi^\prime)^2\over 3p}+{H^2\over A}\right).
\label{ueqp}
\ee
First, let us consider the $S^1/\Bbb{Z}_2$ orbifolded model, with 
a brane at each
end, where the warp factors at these branes are $A_i,\,i=1,2$, and 
the brane tensions are $\sigma_i$. Note that it is possible for
both brane tensions to be positive, if one of them is
located at $y>y_0$ and the other at $y<y_0$. We define the quantities
$Q_1$ and $Q_2$ by
\be
4Q_i^2=\left({\kappa^2\sigma_i\over 3}\right)^2
+{\kappa^2\over 3p}\left({\partial\sigma_i\over
\partial\phi}\right)_i^2.
\ee
Using the jump conditions (\ref{Israel})
along with Eq.\ (\ref{ueqp}), where the constant term in $V(\phi)$
is adjustable (for example, via unimodular gravity), we find that
\baray
k^2&=&{A_1Q_1^2-A_2Q_2^2\over A_1-A_2}
\nonumber\\
H^2&=&{A_1A_2(Q_2^2-Q_1^2)\over A_1-A_2}.
\label{Hp}
\earay
In the regime $A_2 \ll A_1$, these formulae reduce to 
\baray
k^2&\approx&  Q_1^2\nonumber\\
H^2&\approx&  A_2(Q_2^2-Q_1^2),
\label{Hp1}
\earay
where we have assumed that $Q_2^2\sim Q_1^2$.
Note that Eq.\ (\ref{Hp}) requires that
$Q_2^2>Q_1^2$.
The result (\ref{Hp1}) implies that the four dimensional cosmological
constant $\Lambda_4 \propto H^2/A_1$ is suppressed by the ratio $A_2 /
A_1$ of warp factors, which will be vanishingly small for suitable
brane separations. 

Now suppose that our Universe lives on a test
brane located somewhere between the branes 1 and 2.  Denote by $A_U$
the warp factor on our brane, and let brane 1 be the Planck brane.  
Then, we find that the electroweak scale
on our brane is given by $m_{EW}^2=m_P^2A_U/A_1$, where $m_P$ is the
Planck scale. The expansion rate on our brane is
$H_U^2=H^2/A_U$, and therefore
\be
{H_U^2\over m_{EW}^2}={A_1^2A_2(Q_2^2-Q_1^2)\over
m_P^2A_U^2(A_1-A_2)}
\simeq{A_1A_2(Q_2^2-Q_1^2)\over m_P^2A_U^2}.
\ee
Assuming that $Q_i^2\sim m_P^2$, we see that $H_U^2/m_{EW}^2$
can be exponentially small provided that $A_U\gg\sqrt{A_1A_2}$.
Roughly speaking, this will be the case if the test brane on which
our Universe resides is closer to the Planck brane, brane 1,
than it is to brane 2. Turning on a small positive brane tension 
for the visible brane does not change the above qualitative features.
Even for models in which singularities are not absolutely
prevented, they can still be avoided in the orbifold case, which
requires a negative tension brane at one fixed point.
If one only wants to solve
the mass hierarchy problem, this negative tension brane is 
the visible brane\cite{RS1}. If one also wants to solve the
cosmological constant problem, the visible brane must be 
identified with a third brane between the orbifold boundaries,
as in the nonsingular models discussed above. 

This result can be generalized to other $H\neq 0$ spacetimes with
no bulk curvature singularities, and remains valid for uncompactified,
multi-brane models as well. It can also be generalized to cases 
where the branes are charged under some 4-form potentials\cite{FJSTW}.
In this case, the constant term (and/or other coefficients) in 
$V(\phi)$ depends on the 5-form field strengths, and the background
values of these field strengths are determined by satisfying the jump 
conditions at the branes. With the warp factor normalized to unity at 
the Planck brane, $\Lambda_4$ is proportional to $A_p$, the warp factor 
at the nearest particle horizon. This factor can easily be 
exponentially small. Since the visible brane must be closer to 
the Planck brane than the particle horizon, the warp factor $A_v$ at 
the visible brane can easily be exponentially small as well, providing 
a solution to the hierarchy problem\cite{FJSTW}, but $A_v >> A_p$, as 
observed in nature.

\section{Discussion}
\label{discussion}

In this paper, we have constructed several different models
of 5-D brane worlds with bulk scalar fields. 
We gave a general procedure for constructing static ($H=0$)
models with no
curvature singularities in the bulk.
A specific model was constructed corresponding to
compactifying from an 11-D theory with a 7-form field strength down to
5-D, where the 7-form field descends to a 5-form field strength.
By adjusting the 5-form field strength, we were able to render the
model nonsingular. 
In general, we found that nonsingular models are more warped
than the RS model on relatively small distance
scales, but tend towards simple exponential warping on
large scales. Thus, the RS behavior is robust
in the asymptotic regime, although stabilization of branes via
the introduction of bulk scalar fields may require the branes
to reside where the fields produce substantial extra warping
of spacetime.

For expanding ($H \ne 0$) models, we found that 
singularities occur even for those scalar field potentials 
that do not give rise to singularities in the $H=0$ case.
We therefore were compelled to explore a
different method for finding nonsingular models with
$H\neq 0$.  A subclass of these models incorporates a non-singular
surface on which the warp factor $A(y)$ vanishes, analogous to the
particle horizons that occur in the pure AdS case\cite{FJSTW}.
One specific class of such models, which
corresponds to a sinusoidally varying potential in
the nonexpanding case, was treated in some detail.  For
this case we constructed an orbifolded model, with one bounding
brane being the Planck brane, and the other bounding brane (on which
the warp factor is exponentially smaller)
having either positive or negative tension.  In this model,
we live on a test brane between 
the two boundaries branes, and the expansion
rate on our brane is exponentially smaller than 
the local electroweak scale provided that our brane is
located somewhat closer to the Planck brane than to the
other brane.  This feature does not appear to depend in
an essential way on details of our particular model, and
even can be shown to hold when singularities are avoided
(by orbifolding) rather than prevented outright by
choice of potential. Thus, there are models in which the cosmological
constant and mass hierarchy problems can be explained simultaneously.

Two fine-tunings of the parameters of the RS orbifold model
are required to get flat 4-D spacetime\cite{RS1}.
Allowing non-zero $\Lambda_4$ relaxes one of these fine-tunings,
while the other can be eliminated by allowing the bulk cosmological
constant to depend on 5-form field strengths, whose piecewise constant 
values are determined by boundary conditions at the branes. 
When the branes are relatively far apart, $\Lambda_4$ turns out to be 
exponentially small\cite{FJSTW}.  However, the models of Ref.\
\cite{FJSTW} were dynamically unstable.  Bulk scalar fields can be 
introduced to fix brane positions.  When $\Lambda_4=0$, there are two 
sources of fine tuning in models with bulk scalar fields, 
one associated with expressing the
potential $V(\phi)$ in terms of a superpotential, $W(\phi)$
[Eqs.\ (\ref{phipsuper}) and (\ref{vsuper})], and the other
associated with adjusting $W(\phi)$ so that the solution
becomes nonsingular.  Given a potential $V(\phi)$, there is
no guarantee that a $W(\phi)$ can be found that satisfies
Eq.\ (\ref{vsuper}) without tuning the parameters \cite{DeWolfe}.  
For example, a general quartic potential of the form
$V(\phi)=a_0+a_2\phi^2/2+a_4\phi^4/4$ contains three parameters,
whereas Eq.~(\ref{gw1}) or (\ref{gw2}) contains only
two, implying that a relationship among $a_0$, $a_2$ and
$a_4$ is needed in order for $V(\phi)$ to arise from a
superpotential $W(\phi)$.  Moreover, Eq.~(\ref{gw1}) corresponds
to a nonsingular model, whereas Eq.~(\ref{gw2}) is singular;
the two independent parameters in a quartic potential have
to be specifically adjusted to prevent singularities.
The adjustment can be achieved by the introduction of 
adjustable 5-form field strengths, and we may speculate that, 
in an evolving bulk, the 5-form field strengths may relax naturally
(e.g. via bubble nucleation) to values that prevent the occurrence of 
singularities. (A specific example of the role that 5-form
field strengths might play is given in 
Eqs.\ (\ref{comp1})-(\ref{comp4}) of Sec.\ \ref{sec:dr}.)
In general, additional fine-tunings required
to prevent singularities need not be invoked in an
orbifold model, where compactification may simply avoid
the presence of singularities that could otherwise
arise in an uncompactified model.

When $\Lambda_4\neq 0$, some fine tuning
is still needed, apparently, for singularities to be
absent, but we had to resort to a different method,
not based on a superpotential, to construct nonsingular
solutions in this case.  The problem arises because
singularities can only be prevented if $\phi^\prime\to 0$
when $A(y)\to 0$, a condition that does not follow 
readily from a description in terms of a superpotential,
and therefore represents a different kind of tuning.
This suggests that the bulk alone may be supersymmetric,
making a description in terms of a superpotential $W(\phi)$
natural, but that the presence of branes breaks the
supersymmetry completely, and leads to nonzero $\Lambda_4$.
As in the nonexpanding case, it is also possible to
preserve models based on a superpotential without any
special tuning of the parameters in $W(\phi)$ in a
appropriately compactified or orbifolded model, but it
is never possible to relate $V(\phi)$ directly to $W(\phi)$
when $\Lambda_4\neq 0$.

\medskip

We thank Csaba Csaki and Zurab Kakushadze for pointing out earlier 
works that are relevant to this paper.
This research was supported in part by NSF grant PHY-9722189 (E.E.F.),
by other NSF grants (S.-H.H.T.) and by NASA (I.W.).

\end{document}